\documentstyle[11pt]{article}

\columnwidth\textwidth

\title{Power counting degree versus singular  order \\ in the Schwinger model}
\author{Andreas Aste, G\"unter Scharf, Urs Walther
\\Institut f\"ur Theoretische Physik der Universit\"at Z\"urich,
\\ Winterthurerstr. 190, CH-8057 Z\"urich, Switzerland}
\date{October 10, 1997}
\begin{document}
\maketitle
\begin{abstract}  
The importance of a rigorous definition of the singular degree of a
distribution is demonstrated on the case of two-dimensional
QED (Schwinger model). Correct mathematical treatment of second order
vacuum polarization in the perturbative approach is crucial
in order to obtain the Schwinger mass of the photon by resummation.
\end{abstract}
\newpage

\section{Introduction}
The Schwinger model [1] still serves as a very popular laboratory
for quantum field theoretical methods. Although its nonperturbative
properties and their relations to confinement [2,3]
have always been of greatest interest, it is also possible to
discuss the model perturbatively in a straightforward way.
The calculation of the vacuum polarization diagram (VP) at second order
then turns out to be a delicate task, where a careful discussion
of the scaling behaviour of distributions becomes necessary.

We will demonstrate this fact in the framework of causal perturbation theory
in the following.

\section{The causal approach}
In causal perturbation theory, which goes back to a classical paper
by H. Epstein and V. Glaser [4], the $S$-matrix is 
constructed inductively order by order as an operator
valued functional
\begin{equation}
S(g)=1+\sum_{n=1}^\infty{1\over n!}\int d^4x_1\ldots d^4x_n\,
T_n(x_1,\ldots x_n)g(x_1)\ldots g(x_n),
\end{equation}
where $g(x)$ is a tempered test function that switches the interaction.
The first order (e.g. for QED)
\begin{equation}
T_1 (x) = ie:\bar{\Psi}(x) \gamma^\mu \Psi(x): A_\mu(x)
\end{equation}
must be given in terms of the asymptotic free fields. It is a striking
property of the causal approach that {\em no ultraviolet divergences}
appear, i.e. the $T_n$'s are finite and well defined.
The only remnant of the ordinary renormalization theory is a
non-uniqueness of the $T_n$'s due to {\em {finite}} normalization terms.
The adiabatic limit $g(x) \rightarrow 1$ has been 
shown to exist in purely massive theories at each order [4].

To calculate the second order distribution $T_2$, one proceeds as follows:
First one constructs the distribution $D_2(x,y)$
\begin{equation}
D_2 (x,y) = [T_1(x),T_1(y)] \quad ,
\end{equation}
\begin{equation}
{\mbox {supp}} \, D_2 =\{(x-y)\>|\>(x-y)^2\ge 0\} \quad,
\end{equation}
which has causal support.
Then $D_2$ is split into a retarded and an advanced part $D_2=R_2-A_2$,
with
\begin{equation}
{\mbox {supp}} \, R_2 =\{(x-y)\>|\>(x-y)^2\ge 0,\> (x^0-y^0)\ge 0\} \quad,
\end{equation}
\begin{equation}
{\mbox {supp}} \, A_2 =\{(x-y)\>|\>(x-y)^2\ge 0,\> -(x^0-y^0)\ge 0\} \quad.
\end{equation}
Finally $T_2$ is given by
\begin{equation}
T_2(x,y) = R_2(x,y) + T_1(y)T_1(x) = A_2(x,y) - T_1(x)T_1(y) \quad .
\end{equation}

For the massive Schwinger model with fermion mass m,
the part in the Wick ordered distribution
$D_2$ corresponding to VP
\begin{equation}
D_2(x,y) = e^2 [ d_2^{\mu \nu} (x-y)-d_2^{\nu \mu}(y-x) ]
:A_\mu(x) A_\nu (y): + ...
\end{equation}
then becomes
\begin{displaymath}
{\hat d}_2^{\mu \nu} (k) = \frac{1}{2 \pi} \int d^2z \,
d_2^{\mu \nu} (z) e^{ikz}
\end{displaymath}
\begin{equation}
{\hat d}_2^{\mu \nu} (k) = \Bigl(g_{\mu \nu} - \frac{k_\mu k_\nu}{k^2}\Bigr) 
\frac{4m^2}{2 \pi} \frac{1}{k^2 \sqrt{1-4m^2/k^2}} {\mbox {sgn}}(k^0) \Theta(k^2-4m^2) \quad . \label{ip}
\end{equation}


\section{Power counting degree and singular order}

Obviously, ${\hat d}_2^{\mu \nu}$ has power counting degree
$\omega_p = -2$ [6]. But the singular order of the distribution is
$\omega=0$. To show what this means we recall the following
definitions [5,8]:
\vskip 0.4 cm
{\bf Definition 1a:} The distribution ${\hat d}(p) \in {\cal S}'({\bf R}^n)$ has
quasi-asymptotics ${\hat d}_0(p) \not\equiv 0$
at $p=\infty$ with respect to {\em a positive
continuous function} $\rho(\delta), \delta>0$, if the limit
\begin{equation}
\lim_{\delta \rightarrow 0} \rho(\delta) \langle {\hat d} \Bigl( \frac{p}{\delta}
\Bigr) , {\check \varphi}(p) \rangle  = \langle {\hat d}_0 , {\check \varphi}
\rangle
\end{equation}
exists for all ${\check \varphi} \in {\cal S}({\bf R}^n)$.
The Fourier transform of a test function $\varphi(x)$ is defined by
\begin{equation}
{\hat \varphi}(p)=(2 \pi)^{-n/2} \int d^nx \, \varphi (x)
\, e^{ipx}
\end{equation}
By scaling transformation one derives
\begin{equation}
\lim_{\delta \rightarrow 0} \frac{\rho (a\delta)}{\rho(\delta)}=a^\omega
\equiv \rho_0(\delta) \label{scala}
\end{equation}
with some real $\omega$. Thus we call $\rho (\delta)$ the power-counting function.
The equivalent definition in x-space reads as follows:
\vskip 0.4 cm
{\bf Definition 1b:} The distribution $d(x) \in {\cal S}'(\bf{R}^n)$ has 
{\em quasi-asymptotics} $d_0 (x) \not\equiv 0$ at $x=0$ with respect to 
{\em a positive continuous function} $\rho (\delta)$, $\delta>0$, if the limit 
\begin{equation}
\lim_{\delta \rightarrow 0} \rho(\delta)\delta^n d(\delta x) = d_0 (x)
\end{equation}
exists in ${\cal S}' (\bf{R}^n)$.

\vskip 0.4 cm
{\bf Definition 2:} The distribution $d(x) \in {\cal S}'({\bf R}^n)$ is called
{\em singular of order $\omega$}, if it has quasi-asymptotics $d_0(x)$
at $x=0$, or its Fourier transform has quasi-asymptotics ${\hat d}_0(p)$ at
$p=\infty$, respectively, with power-counting function $\rho(\delta)$
satisfying
\begin{equation}
\lim_{\delta \rightarrow 0} \frac{\rho(a\delta)}{\rho(\delta)}=a^\omega \quad
\forall a>0.
\end{equation}
Equation (\ref{scala}) implies
$$
a^n \langle {\hat d}_0(p),{\check \varphi}(ap) \rangle = \langle
{\hat d}_0
\Bigl(\frac{p}{a}\Bigr),{\check \varphi}(p)\rangle=a^{-\omega}\langle {\hat d}_0(p),{\check \varphi}(p) \rangle
$$
\begin{equation}
=\langle d_0(x),\varphi\Bigl(\frac{x}{a}\Bigr)\rangle=a^n
\langle d_0(ax),\varphi (x)\rangle=a^{-\omega} \langle d_0(x),\varphi(x)
\rangle,
\end{equation}
i.e. ${\hat d}_0$ is homogeneous of degree $\omega$:
\begin{equation}
{\hat d}_0 \Bigl( \frac{p}{a}) = a^{-\omega} {\hat d}_0(p),
\end{equation}
\begin{equation}
d_0(ax)=a^{-(n+\omega)} d_0(x).
\end{equation}
This implies that $d_0$ has power-counting function
$\rho(\delta)=\delta^\omega$ and
the singular order $\omega$, too. In particular, we have the following
estimates for $\rho(\delta)$ [5]: If $\epsilon>0$ is an arbitrarily
small number, then there exist constants $C,C'$ and $\delta_0$ such that
\begin{equation}
C \delta^{\omega+\epsilon} \geq \rho(\delta) \geq C' \delta^{\omega-\epsilon}
\quad , \quad \delta<\delta_0.
\end{equation}

Applying the above definitions to ${\hat d}_2^{\mu \nu} (k)$, we obtain after a
short calculation the quasi-asymptotics
\begin{equation}
\lim_{\delta \rightarrow 0} {\hat d}_2^{\mu \nu}(k/\delta) =
\frac{1}{2 \pi} \Bigl(g^{\mu \nu} k^2 - k^\mu k^\nu \Bigr)
\delta(k^2) {\mbox {sgn}}(k^0) \quad , \label{res}
\end{equation}
and we have $\rho(\delta)=1$, hence $\omega=0$. Note that the
$g^{\mu \nu}$-term in (\ref{res}) does not contribute to the quasi-asymptotics.
The reason for the result (\ref{res}) can be explained by the existence of a
{\em {sum rule}} [7]
\begin{equation}
\int \limits_{4m^2\delta^2}^{\infty} d(q^2) \frac{\delta^2 m^2}
{q^4 \sqrt{1-\frac{4m^2\delta^2}{q^2}}} = \frac{1}{2} \quad , \label{asy}
\end{equation}
so that the l.h.s. of (\ref{res}) is weakly
convergent to the r.h.s.
In spite of ${\mbox{sgn}} (k^0)$,
the r.h.s. of (\ref{res}) is a well-defined tempered distribution
due to the factor $(g^{\mu \nu} k^2 - k^\mu k^\nu)$.

This has the following consequence: The retarded part $r_2^{\mu \nu}$ of
$d_2^{\mu \nu}$ would be given in the case $\omega < 0$ by the unsubtracted splitting formula
\begin{displaymath}
r_2^{\mu \nu}(k) = \frac{i}{2 \pi} \int \limits_{-\infty}^{\infty}
\frac{dt}{1-t+i0}d_2^{\mu \nu}(tk) 
\end{displaymath}
\begin{equation}
=\frac{im^2}{\pi^2} \Bigl( g^{\mu \nu} - \frac{k^\mu k^\nu}{k^2} \Bigr)
\frac{1}{k^2 \sqrt{1-4m^2/k^2}} \log \frac{\sqrt{1-4m^2/k^2}+1}
{\sqrt{1-4m^2/k^2}-1} \quad ,
\quad k^2>4m^2,k^0>0.
\end{equation}
This distribution will vanish in the limit $m \rightarrow 0$, and
the photon would remain massless. But since we have $\omega=0$,
the subtracted splitting formula [5] {\em {must}} be used:
\begin{displaymath}
r_2^{\mu \nu}(k) = \frac{i}{2 \pi} \int \limits_{-\infty}^{\infty}
\frac{dt}{(t-i0)^{\omega+1}(1-t+i0)}d_2^{\mu \nu}(tk)
\end{displaymath}
\begin{equation}
=\frac{im^2}{\pi^2} \Bigl( g^{\mu \nu} - \frac{k^\mu k^\nu}{k^2} \Bigr)
\Bigl( \frac{1}{k^2 \sqrt{1-4m^2/k^2}} \log \frac{\sqrt{1-4m^2/k^2}+1}
{\sqrt{1-4m^2/k^2}-1} + \frac{1}{2m^2} \Bigr) \quad ,
\quad k^2>4m^2,k^0>0 \, .
\end{equation}
The new last term survives in the limit $m \rightarrow 0$. After resummation
of the VP bubbles it gives the well-known Schwinger mass
$m_s^2=e^2/\pi$ of the photon.
Consequently, the difference between simple power-counting and the
correct determination of the singular order is by no means a mathematical
detail, it is terribly important for the physics.
\vskip 1cm
{\it References}\vskip 1cm
\begin{enumerate}
\item Schwinger, J.: Gauge invariance and mass II, {\it Phys. Rev.}
 {\bf 128} (1962), 2425-2429.

\item Casher, A., Kogut, J. and Susskind, L.: Vacuum Polarization and the
Quark-Parton Puzzle, {\it Phys. Rev. Lett.} {\bf 31} (1973), 792-795.

\item Casher, A., Kogut, J. and Susskind, L.: Vacuum polarization and the absence of free quarks, {\it Phys. Rev. D} {\bf 10} (1974), 732-745.

\item Epstein, H. and Glaser, V.: The Role of Locality in Perturbation Theory,
{\it Ann.Inst.Poincar\'e A} {\bf 29} (1973), 211-295.

\item Scharf, G.: {\it Finite Quantum Electrodynamics: the causal approach},
second edition, 1995 Springer Verlag, 1995.

\item Weinberg, S.: High-energy Behaviour in Quantum Field Theory, {\it Phys. Rev.} {\bf 118} (1960), 838-849.

\item Adam, C., Bertlmann, R. A. and Hofer, P.: Dispersion Relation
Approach to the Anomaly in 2 Dimensions, {\it Z. Phys. C} {\bf 56}
(1992) 123-127.

\item Vladimirov, V. S., Drozzinov, Y. N. and Zavialov, B. I.: {\it Tauberian
Theorems for Generalized Functions}, Kluwer Acad. Publ., 1988.
\end{enumerate}

%
%
%
%
%
%
%
%

\end{document}